\documentstyle[11pt,newpasp,twoside]{article}
\input{epsf}
\reversemarginpar

\newcommand{\EQ}{\begin{equation}}
\newcommand{\EN}{\end{equation}}
\newcommand{\EQA}{\begin{eqnarray}}
\newcommand{\ENA}{\end{eqnarray}}
\newcommand{\eq}[1]{(\ref{#1})}
\newcommand{\Eq}[1]{Eq.~(\ref{#1})}

\newcommand{\Fig}[1]{Fig.~\ref{#1}}

\newcommand{\bra}[1]{\langle #1\rangle}

\newcommand{\meanBB}{\overline{\mbox{\boldmath $B$}}}
\newcommand{\meanJJ}{\overline{\mbox{\boldmath $J$}}}

\newcommand{\uu}{\mbox{\boldmath $u$} {}}

\newcommand{\bb}{\mbox{\boldmath $b$} {}}

\newcommand{\BB}{\mbox{\boldmath $B$} {}}

\newcommand{\AAA}{\mbox{\boldmath $A$} {}}

\newcommand{\jj}{\mbox{\boldmath $j$} {}}

\newcommand{\oo}{\mbox{\boldmath $\omega$} {}}

\newcommand{\dd}{{\rm d} {}}

\newcommand{\ea}{{\rm et al. }}

\newcommand{\yapj}[3]{ #1, {ApJ }{#2}, #3}
\newcommand{\yapjl}[3]{ #1, {ApJ (Letters) }{#2}, #3}

\newcommand{\yan}[3]{ #1, {Astr. Nachr. }{#2}, #3}
\newcommand{\yana}[3]{ #1, {A\&A }{#2}, #3}

\newcommand{\yjfm}[3]{ #1, {JFM }{#2}, #3}

\newcommand{\ysov}[3]{ #1, {Sov. Astron. }{#2}, #3}

\newcommand{\ysph}[3]{ #1, {Solar Phys. } {#2}, #3}

\newcommand{\yjour}[4]{ #1, {#2} {#3}, #4}
\newcommand{\ybook}[3]{ #1, {#2} (#3)}

\newcommand{\sapj}[1]{ #1, {ApJ } (submitted)}
\newcommand{\sana}[1]{ #1, {A\&A } (submitted)}

\begin{document}
\title{The solar dynamo: old, recent, and new problems}
\author{Axel Brandenburg}
\affil{NORDITA, Blegdamsvej 17, DK-2100 Copenhagen \O, Denmark; and\\
Department of Mathematics, University of Newcastle upon Tyne, NE1 7RU, UK}

\begin{abstract}
A number of problems of solar and stellar dynamo theory are briefly
reviewed and the current status of possible solutions is discussed.
Results of direct numerical simulations are described in view of
mean-field dynamo theory and the relation between the $\alpha$-effect and
the inverse cascade of magnetic helicity is highlighted. The possibility of
`catastrophic' quenching of the $\alpha$-effect is explained in terms
of the constraint placed by the conservation of magnetic helicity.
\end{abstract}

\section{Introduction}

The solar dynamo problem has always been plagued by a number of
problems. One of the most outstanding is the question of why
the solar dynamo wave travels toward the equator and not the other
way around. According to $\alpha\Omega$-dynamo theory the dynamo wave
travels equatorward only if the sign of the product of alpha-effect
and radial angular velocity gradient, $\partial\Omega/\partial r$, is
{\it negative}. However, cyclonic convection gives rise to a positive
$\alpha$-effect (Parker 1955, Steenbeck, Krause, \& R\"adler 1966), so one
needs a negative sign of $\partial\Omega/\partial r$. Thus, a problem has
arisen since the mid-eighties (Parker 1987), i.e.\ since helioseismology
took away the freedom of dynamo theorists to adopt a suitable rotation
profile. This left only the choice of adopting an appropriate profile
(and sign!) for the alpha-effect. Not surprisingly, dynamo theorists
were labelled as being able to reproduce almost anything, but in reality
they were not even able to do that, because there was one more problem:
the phase relation between poloidal and toroidal fields (Stix 1976).
The sign of the mean radial magnetic field, ${\overline B}_r$, is
observed to be almost always opposite to the sign of the mean toroidal field,
${\overline B}_\phi$. The phase relation between ${\overline B}_r$ and
${\overline B}_\phi$ is determined directly by $\partial\Omega/\partial
r$, because the shear turns radial field into toroidal. This problem
is actually rather general in its nature and quite independent of
dynamo theory.

One possible solution to this dynamo dilemma (Parker 1987) may lie in
the possibility that the sense of the dynamo wave can be determined
by the sense of the meridional circulation: a poleward circulation in
the upper layers corresponds to equatorward motion in the lower layers,
and if the sunspot activity is governed by fields at the bottom of the
convection zone this circulation may control the migration of activity
regions Durney 1996, Choudhuri \ea 1996). A systematic survey of solutions
with meridional circulation (K\"uker \ea 2000) has revealed that the sense
of the dynamo wave is reversed only if the following conditions are met:
the magnetic eddy diffusivity is small enough, the circulation is strong
enough, and the $\alpha$-effect is sufficiently supercritical. If this is
the case, K\"uker \ea (2000) also find ${\overline B}_r$ and ${\overline
B}_\phi$ to be approximately in antiphase, as observed.

Over the past ten years some rather more fundamental problems have
shadowed the attempts to model closely the solar field geometry: does the
$\alpha$-effect really work in high Reynolds number flows like the sun,
or is it totally suppressed by the magnetic field?

\section{The helicity constraint}

A serious problem that has only recently received attention results from
magnetic helicity conservation. The problem is relatively easily explained
and applies to all fields that have magnetic helicity. In particular,
it applies to the large scale magnetic fields generated by $\alpha^2$
or $\alpha\Omega$ dynamos, but not, for example, to fields generated by
a small scale dynamo which may produce significant helicity fluctuations,
but no net magnetic helicity.

Magnetic helicity can only change resistively or through flux on the
boundaries. Thus, a large scale field with magnetic helicity can only
be generated on a resistive timescale or, if there are suitable losses
of the right kind on the boundaries, field generation can proceed on
a dynamical timescale. The latter alternative is not straightforward,
because such losses always involve losses of magnetic energy too. This
latter alternative will be discussed later. First, however, we shall look
more quantitatively at the consequences of resistively slow changes of
helicity in closed or periodic boxes.

The condition of helicity conservation together with the assumption
that the large scale field is helical result in a simple condition
for the energy of the mean magnetic field (Brandenburg 2000, hereafter
referred to as B2000)
\EQ
\bra{\meanBB^2}/B_{\rm eq}^2\approx k_{\rm f}\ell_{AB}\left[1-
\exp(-2\eta k_1^2 (t-t_{\rm sat})\right],
\label{helconstr}
\EN
where $k_{\rm f}$ is the wavenumber of the (kinetic) energy carrying
eddies (e.g.\ the forcing wavenumber), $k_1$ is the smallest possible
wavenumber, $\eta$ is the {\it microscopic} magnetic diffusivity, $t$ is
time, $t_{\rm sat}$ is the time when the field at small and intermediate
scales saturates, and $\ell_{AB}$ quantifies the degree to with the large
scale field is helical. This relation is rather general and independent
of the actual model of field amplification. If the field is not fully
helical, then $\ell_{AB}$ will be reduced, but the important point
here is that full saturation is only obtained after a large scale ohmic
diffusion time, $1/\eta k_1^2$.

\begin{figure}[h!]\epsfbox{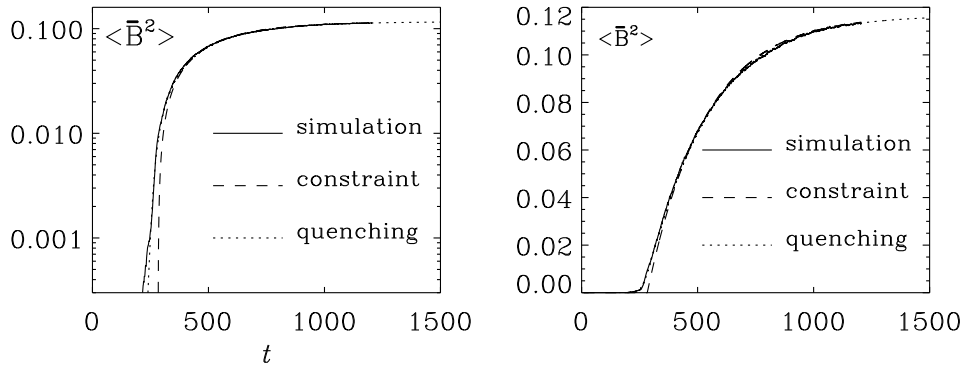}\caption[]{
Evolution of $\bra{\meanBB^2}$ for Run~3 of B2000, compared with the constraint
imposed by helicity conservation (dashed lines). Also plotted is the
result of a mean-field dynamo with `catastrophic' alpha and eta quenching
(dotted lines).
}\label{pjbm_decay_nfit_run3}\end{figure}

The important point is that all box simulations with helicity have to
obey this behavior. This is shown in \Fig{pjbm_decay_nfit_run3} for one
particular example, but this constraint is indeed very well obeyed
by all other simulations presented in B2000.

\epsfxsize=13cm

The helicity constraint has two properties: full saturation of the large scale
field (as opposed to the small scale field) happens on a resistive time scale,
$1/\eta k_1^2$, and secondly, the energy of the maximum field strength
in units of the equipartition field strength is given by the
product $k_{\rm f}\ell_{AB}$, and for maximally helical fields we have
$\ell_{AB}=k_1^{-1}$. In our particular runs we have $k_{\rm f}/k_1=5$,
so we can reach maximally five times supercritical energies of the large
scale field. (In B2000 there is also run with $k_{\rm f}/k_1=30$, which
does indeed give correspondingly higher field strengths.) The reason we
have this factor 5 here can be motivated as
follows. The large scale dynamo is driven by some effective $\alpha$ which
has contributions from the kinetic helicity and the current helicity of
the small scale field which combine to a residual helicity (Pouquet \ea
1976, Field \ea 1999). Near equipartition the residual helicity is very
small, so kinetic helicity and small scale current helicity nearly balance
each other. The kinetic helicity, $\bra{\oo\cdot\uu}$, is proportional to
$k_{\rm f}\bra{\uu^2}$, so we know how big that is. On the other hand,
the small scale current helicity, $\bra{\jj\cdot\bb}$, is exactly
the term which drives magnetic helicity, $\bra{\AAA\cdot\BB}$. In
the steady state $\bra{\AAA\cdot\BB}$ can no longer change, so
$\bra{\jj\cdot\bb}$ must be balanced with the contribution from large
scales, $\bra{\meanJJ\cdot\meanBB}$, which determine then the large
scale field strength, $\bra{\meanJJ\cdot\meanBB}=k_1\bra{\meanBB^2}$.
This is -- in words -- the reason why superequipartition field strengths
can be attained. (For mathematical details see B2000.)

\subsection{Catastrophic quenching}

Exactly the same resistive behavior as in \Eq{helconstr} can be {\it
reproduced} using the $\alpha^2$ dynamo equation (e.g.\ Moffatt 1978)
with simultaneous quenching of $\alpha$ and the turbulent magnetic
diffusivity, $\eta_{\rm t}$, in the form
\EQ
\alpha={\alpha_0\over1+\alpha_B\meanBB^2\!/B_{\rm eq}^2},\quad
\eta_{\rm t}={\eta_{\rm t0}\over1+\eta_B\meanBB^2\!/B_{\rm eq}^2},
\label{quench_both}
\EN
where we assume $\alpha_B=\eta_B$. We focus here on the case of
homogeneous turbulence in which case the large scale magnetic field
is a force-free field whose energy density is very nearly uniform, so
$\meanBB^2$ is just a function of time and we can obtain the solution
$\meanBB=\meanBB(t)$ in closed form (B2000).

A natural by-product of this model is that the quenching coefficient
$\alpha_B$ (which is assumed to be the same as $\eta_B$) must be
proportional to the magnetic Reynolds number! This type of quenching
was suggested earlier on phenomenological grounds by Cattaneo \&
Vainshtein (1991) and Vainshtein \& Cattaneo (1992), but the conclusion
was thought to be that for dynamically important field strengths
($|\meanBB|\rightarrow B_{\rm eq}$) the two turbulent transport
coefficients would be so strongly quenched that such fields cannot be
the result of a mean-field dynamo. That was the reason that such quenching
was referred to as `catastrophic'. We now see however that there is
nothing catastrophic about it and that field strength even in {\it excess}
of $B_{\rm eq}$ can be generated.

The significance of these results is that they provide an excellent fit to
the numerical simulations; see \Fig{pjbm_decay_nfit_run3} where we present
the evolution of $\bra{\meanBB^2}$ for Run~3 of B2000. The
dynamo equations with appropriate quenching expressions
can therefore be used to extrapolate to astrophysical conditions. The time
$\tau_{\rm eq}$ required to convert the small
scale field generated by the small scale dynamo to a large scale field
increases linear with the magnetic Reynolds number,
$R_{\rm m}$. Apart from some coefficients of order unity the ratio of
$\tau_{\rm eq}$ to the turnover time is therefore just $R_{\rm m}$. For
the sun this ratio would be $10^8-10^{10}$. However, before interpreting
this result further one really has to know whether or not the presence
of open boundary conditions could alleviate the issue of very long
timescales for the mean magnetic field. Furthermore, it is not clear
whether the long timescales discussed above have any bearing on the
cycle period in the case of oscillatory solutions. The reason this is
not so clear is because for a cyclic dynamo the magnetic helicity in
each hemisphere stays always of the same sign and is only slightly
modulated. It is likely that this modulation pattern is advected
precisely with the meridional circulation, in which case the helicity
could be nearly perfectly conserved in a lagrangian frame. This would
support the suggestion of Durney (1995) and Choudhuri, Sch\"ussler, \&
Dikpati (1995) that the dynamo wave travels mainly because of meridional
circulation.

\section{Quadratic or cubic quenching?}

When the first nonlinear mean-field dynamo models with $\alpha$-quenching
were calculated numerically (e.g., Jepps 1975, Ivanova \& Ruzmaikin 1977),
a simple quenching formula of the form $\alpha\sim1/(1+B^2)$ was used.
This was really nothing more than just a simple fix to the problem that
a quadratic formula of the form $\alpha\sim1-B^2$ may overshoot and change
sign. We emphasize that
a quenching formula of the form $\alpha\sim1/(1+B^2)$ was never obtained
rigorously. Instead, the correct quenching behavior, at least in the low
magnetic Reynolds number limit, had cubic limiting behavior (Moffatt 1972,
R\"udiger 1974).

\begin{figure}[h!]\epsfbox{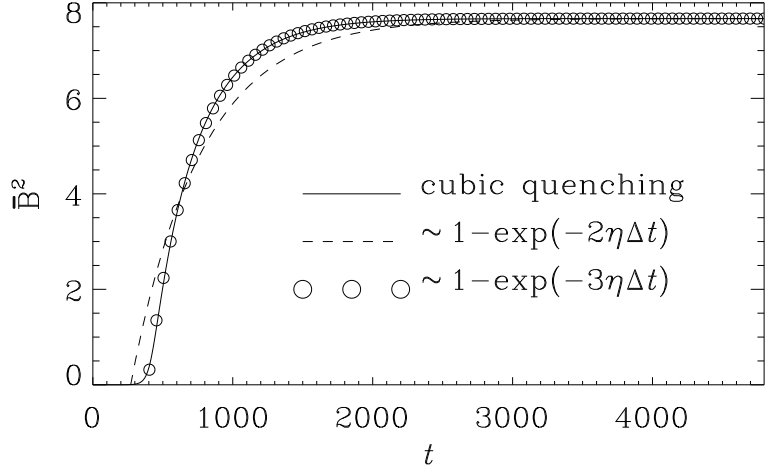}\caption[]{
Clear disagreement between mean-field model with cubic quenching formula (solid line)
and the result from the helicity constraint (dashed line). For comparison the helicity
constraint formula with a $3\eta$ factor (instead of $2\eta$) is also shown as
open symbols.
}\label{pp_cubic}\end{figure}

We are now for the first time in a position to check whether the limiting
behavior for higher magnetic Reynolds numbers is quadratic or cubic.
In \Fig{pp_cubic} we plot the evolution of the magnetic energy for a
model with cubic $\alpha$-quenching,
\EQ
\alpha=\alpha_0/[1+(\alpha_B\meanBB^2/B_{\rm eq}^2)^n],\quad n=3/2.
\EN
It turns out that in this case the evolution of $\meanBB^2$ does not
agree with the limiting behavior enforced by the helicity constraint
\eq{helconstr}; see \Fig{pp_cubic}. This confirms that the phenomenological
expression \eq{quench_both} is actually correct, at least for the
homogeneous $\alpha^2$ dynamo. (We shall see later that for inhomogeneous
dynamos with shear the tensorial nature of $\alpha$-effect and turbulent
diffusivity cannot be ignored, and that the quenching works differently for
different components.)

\epsfxsize=12cm

One might think that a catastrophic quenching expression of
the form $1-\alpha_B\meanBB^2\!/B_{\rm eq}^2$, as opposed to
$1/(1+\alpha_B\meanBB^2\!/B_{\rm eq}^2)$, for both $\alpha$ and
$\eta_{\rm t0}$ may also reproduce the resistively dominated growth of
the field, but this is not the case. To show this we consider the steady
state of the $\alpha^2$-dynamo in the one-mode approximation, so
\EQ
{\dd\ln\meanBB^2\over\dd t}=0=(\alpha_0 k-\eta_{\rm t0}k^2)
(1-\alpha_B\meanBB^2\!/B_{\rm eq}^2)-\eta k^2,
\EN
which leads to the final field strength $B_{\rm fin}$ given by
\EQ
\alpha_B B_{\rm fin}^2/B_{\rm eq}^2
=1-{\eta k^2\over\alpha_0 k-\eta_{\rm t0}k^2}
={\alpha_0 k-\eta_{\rm T0}k^2\over\alpha_0 k-\eta_{\rm t0}k^2}\approx1,
\label{multquench}
\EN
where $\eta_{\rm T0}=\eta+\eta_{\rm t0}$ is the total (microscopic plus
turbulent) magnetic diffusivity. Since we know from the simulations
that $B_{\rm fin}/B_{\rm eq}={\cal O}(1)$, \Eq{multquench}
implies that $\alpha_B={\cal O}(1)$, which would then not reproduce the
resistively dominated growth.

In order to appreciate the difference to a quenching formula of the form
$1/(1+\alpha_B\meanBB^2\!/B_{\rm eq}^2)$, we write down the corresponding
equation for that case:
\EQ
{\dd\ln\meanBB^2\over\dd t}=0={\alpha_0 k-\eta_{\rm t0}k^2\over
1+\alpha_B\meanBB^2\!/B_{\rm eq}^2}-\eta k^2,
\EN
so
\EQ
0=\alpha_0 k-\eta_{\rm t0}k^2-\eta k^2(1+\alpha_B\meanBB^2\!/B_{\rm eq}^2)
=\alpha_0 k-\eta_{\rm T0}k^2-\eta k^2\alpha_B\meanBB^2\!/B_{\rm eq}^2,
\EN
which leads to a final field strength $B_{\rm fin}$ given by
\EQ
\alpha_B B_{\rm fin}^2/B_{\rm eq}^2
={\alpha_0 k-\eta_{\rm T0}k^2\over\eta k^2}\gg1,
\EN
so $B_{\rm fin}/B_{\rm eq}={\cal O}(1)$ implies $\alpha_B\gg1$, as in
\Eq{quench_both}, which we know leads to resistively dominated growth.

Although we cannot exclude that there may be other
possible quenching formulae than \Eq{quench_both}, it is clear that
it is easy to come up with other plausible choices which do not obey
the helicity constraint \eq{helconstr}. However, despite its success,
\eq{helconstr} cannot be regarded as fundamental in that under other
circumstances it may not describe the $\alpha$-effect correctly. In
the following we describe first an example where \Eq{quench_both} gives
a good prediction of what is then also verified in simulations, and then
we turn to another example where \Eq{quench_both} does poorly.

\section{Open boundaries}

\begin{figure}[h!]\epsfbox{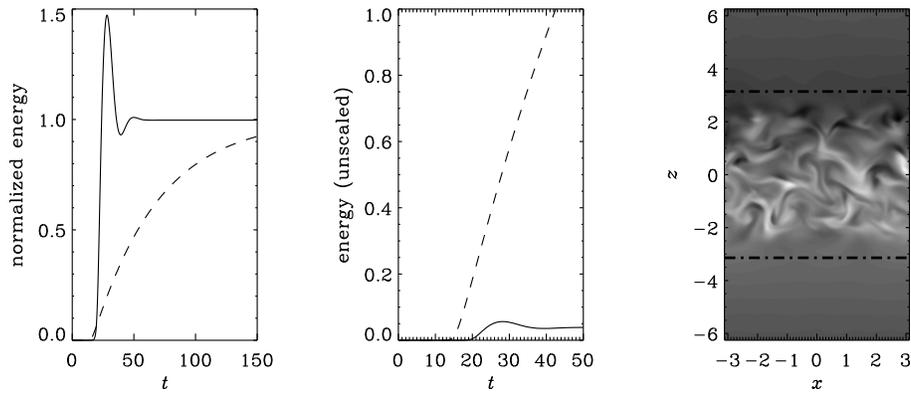}\caption[]{
Evolution of magnetic energy (first panel: normalized to the final
value, second panel: unscaled) for a mean-field model with catastrophic
quenching and either open boundaries (solid lines) or periodic boundaries
(dashed lines).  In the third panel we show a meridional cross-section
of the toroidal magnetic field from a simulation with open boundaries
in the $z$-direction at $\pm\pi$. Different shades represent different
values of the toroidal field strength. The parameters of the mean-field
model are $\alpha_0=2$, $\eta_{\rm t0}=1$, $\eta=0.01$, $k=1$, so
the kinematic growth rate is $\lambda=\alpha_0 k-\eta_{\rm t0}k^2=1$,
and with $k_{\rm f}=5$ we have $\alpha_B=\lambda/(\eta k k_{\rm f})=20$.
In the non-periodic case a vertical field boundary condition was adopted.
}\label{pgrowth+image}\end{figure}

The main problem of why the saturation of the large scale dynamo
progresses on a resistive timescale is, mathematically speaking, that
there is no other term on the right hand side of the magnetic helicity
equation. However, this changes if the volume under consideration is no
longer periodic or closed, so that there can be a transport of magnetic
helicity through the boundaries. However, what tends to happen is that
the loss of helicity occurs mostly on large scales. Associated with
this is a significant loss of large scale magnetic energy. Thus, although
it is true that the time scale gets reduced, the saturation energy
decreases, making it harder to reach significant field strengths of
the large scale field. (The small scale field is unaffected by this.)
In fact, looking at the field evolution on an absolute scale shows
that the loss of magnetic helicity and energy simply terminates the
growth at a lower level; see \Fig{pgrowth+image}.

\epsfxsize=13cm

One may speculate that lower saturation levels are still acceptable
in many astrophysical settings, because if the dynamo is sufficiently
supercritical (like in accretion discs and many stars, but possibly not
in galaxies), the tendency to reduce the saturation level could be offset
by a correspondingly stronger dynamo.

\section{The effects of shear}

\begin{figure}[h!]\epsfbox{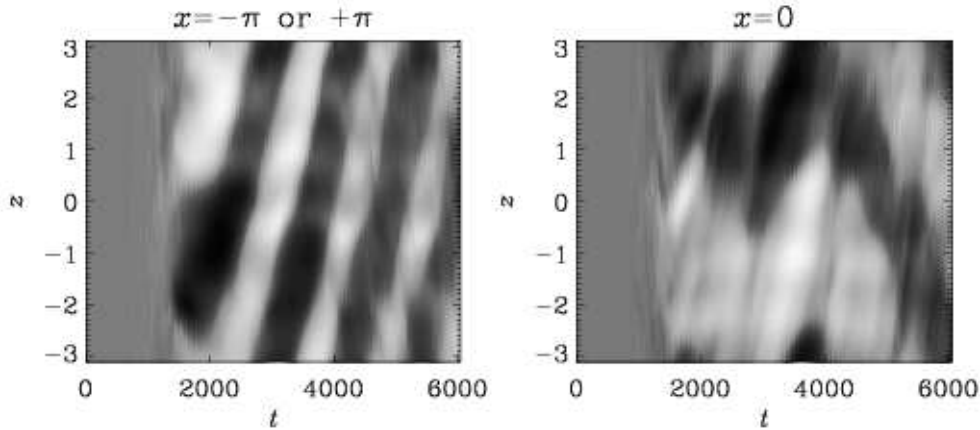}\caption[]{
Butterfly (space-time) diagram of the toroidal mean field at two different
$x$-positions where the sign of the shear is either negative or positive,
giving rise to dynamo waves traveling in opposite direction.
}\label{pbutter_sol}\end{figure}

The presence of shear is interesting because in that case oscillatory
solutions can be expected. This is indeed what happens; see the
butterfly diagram in \Fig{pbutter_sol}. In this model the oscillation
frequency is $2\pi/1000\approx0.006$ (in nondimensional units). This
value lies just between the ohmic and dynamic rates of change, $\eta
k_1^2=5\times10^{-4}$ and 0.08, respectively. This suggests that the
cycle period is not affected by the magnetic helicity constraint in
the same way as the growth time of the dynamo, which is a resistive
timescale. For a full account of this work see Brandenburg \ea (2000).

\epsfxsize=13cm

\section{Conclusions}

An important step has been reached in modelling dynamo action in
astrophysics in that we are now beginning to bridge the gap between
direct three-dimensional simulations on the one hand and the mean-field
approach on the other. Many of the conclusions reached earlier in the
framework of mean-field theory do apply, but there are also important
restrictions. Most importantly, the alpha and eta coefficients have
to obey some rather general restrictions resulting from helicity
conservation. Further extensions to the theory, for example the inclusion
of open boundaries, can straightforwardly be modelled.  However, it seems
that open boundaries do not enhance dynamo action, as one may have hoped,
but instead they lead to a lower field strength at saturation, which is
now however reached earlier on a dynamical timescale. In the presence
of shear the helicity constraint still applies, but the modelling in
terms of mean-field theory is not yet fully understood because anisotropies
in the quenching become very important.

\end{document}